\documentclass[manuscript]{aastex}

\shorttitle{A Search for Nanosecond Optical Transients from Nearby Stars}
\shortauthors{Hanna {\it et al.}}

\begin{document}

\centerline{\includegraphics[scale=0.70]{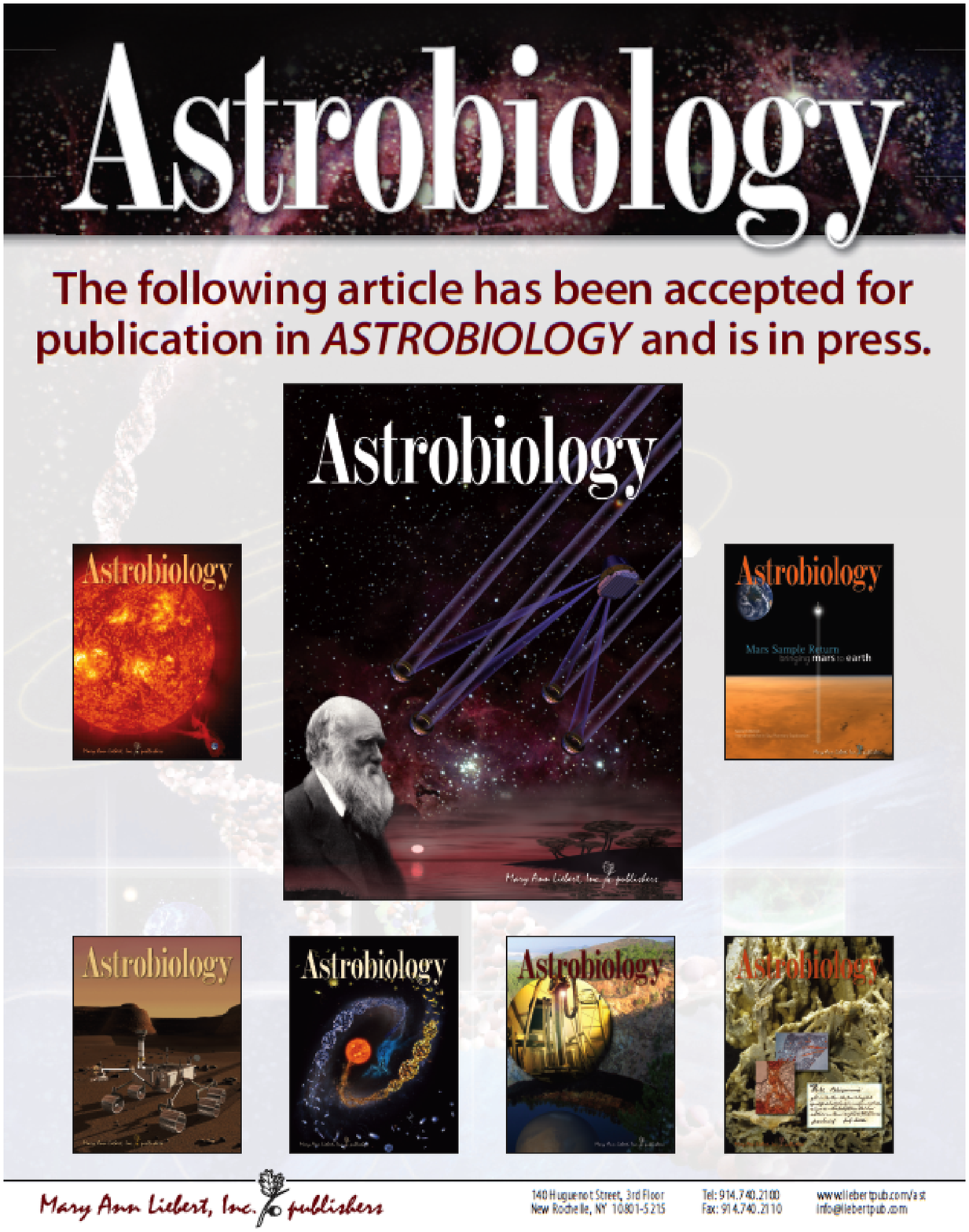}}
\vskip 0.5 cm


\title{OSETI with STACEE: \\ A Search for Nanosecond Optical Transients \\ from Nearby Stars}


\author{D.S. Hanna \altaffilmark{1},
J. Ball \altaffilmark{2,7},
C.E. Covault \altaffilmark{3},
J. E. Carson \altaffilmark{1},
D.D. Driscoll \altaffilmark{3,8},
P. Fortin \altaffilmark{4},
D.M. Gingrich \altaffilmark{5},
A. Jarvis \altaffilmark{2},
J. Kildea \altaffilmark{1,9},
T. Lindner \altaffilmark{1,10},
C. Mueller \altaffilmark{1,11},
R. Mukherjee \altaffilmark{4},
R.A. Ong \altaffilmark{2},
K. Ragan \altaffilmark{1},
D.A. Williams \altaffilmark{6}, and
J. Zweerink \altaffilmark{2}}

\altaffiltext{1}{Physics Department, McGill University,
Montreal, QC H3A 2T8}

\altaffiltext{2}{Department of Physics and Astronomy, University of California,
  Los Angeles, CA 90025} 

\altaffiltext{3}{Department of Physics, Case Western Reserve University,
Cleveland, OH 44106, USA}

\altaffiltext{4}{Department of Physics and Astronomy, Barnard College, 
Columbia University, New York, NY 10027}

\altaffiltext{5}{Department of Physics, University of Alberta,  Edmonton, AB 
T6G2G7 and  TRIUMF, Vancouver, BC V6T 2A3}

\altaffiltext{6}{Santa Cruz Institute for Particle Physics, University of
  California, Santa Cruz, CA 95064}

\altaffiltext{7}{
present address: Gemini Observatory, Hilo, HI 96720}

\altaffiltext{8}{
present address: Kent State University, Ashtabula Campus, Ashtabula, OH 44004 }

\altaffiltext{9}{
present address: Fred Lawrence Whipple Observatory, 
Harvard-Smithsonian Center for Astrophysics, Amado AZ, 85645}

\altaffiltext{10}{
present address: Department of Physics and Astronomy, University of 
British Columbia, Vancouver, BC V6T 1Z1}

\altaffiltext{11}{
present address: Sander Geophysics Ltd., Ottawa, ON, K1V 1C1}


\begin{abstract}
We have used the STACEE high-energy gamma-ray detector to look for fast 
blue-green laser pulses from the vicinity of 187 stars. 
The STACEE detector offers unprecedented light-collecting capability
for the detection of nanosecond pulses from such lasers. 
We estimate STACEE's sensitivity to be approximately 10 photons/m$^2$
at a wavelength of 420 nm. 
The stars have been chosen because their characteristics are such that they may
harbor habitable planets and they are 
relatively close to Earth.
Each star was observed  for 10 minutes and we found no evidence
for laser pulses in any of the data sets.
\end{abstract}





\section{Introduction}

The search for extra-terrestrial intelligence (SETI) has been ongoing
since the publication of the seminal article by Cocconi and 
Morrison~\citep{cocconi}. 
For many years, searches for signals were performed only
at radio wavelengths despite the fact that searches at optical
wavelengths had been suggested as early as 1961~\citep{townes}, following the
invention of the laser.  However, spectacular improvements in laser
technology over the last few decades have dramatically strengthened
the case for optical searches.  We can now imagine a laser system,
built with presently known technology, that could produce detectable
signals over distances to nearby stars. Arguments based on signal-to noise,
pulse dispersion and energy budget considerations support the idea
that laser pulses may be an effective way to conduct interstellar
communication~\citep{howard}. During the past decade, the field of
`optical search
for extra-terrestrial intelligence' (OSETI) has developed as an
established sub-discipline, and several optical searches have been
completed or are now under way~\citep{howard,bhathal,stone,holder,howard07}.

Here we present the results of a search for nanosecond optical
transients via the Solar Tower Atmospheric Cherenkov Effect
Experiment (STACEE). 
The paper is organized as follows.
We begin with a description of the STACEE detector and observation features
relevant to the search. 
We list the targets observed and the data sets obtained and we describe 
calibration procedures.
Finally, we explain the analysis procedure and present the results of our 
search. 

\section{The STACEE Experiment}

The STACEE experiment
was designed for high energy gamma-ray astronomy in 
the energy range from 100 GeV to 10 TeV. It makes use of the atmospheric 
Cherenkov technique whereby energetic astrophysical gamma rays are detected
and measured via the Cherenkov light generated by the relativistic particles
in the air showers which result when the gamma rays interact
in the upper atmosphere.
The wavelength of the light used for this technique is in the range 
from 300 to 600 nm. 
More details on the history and present state of ground-based gamma-ray
astronomy can be found elsewhere~\citep{weekes}.

\begin{figure}
\centerline{\includegraphics[width=1.0\textwidth]{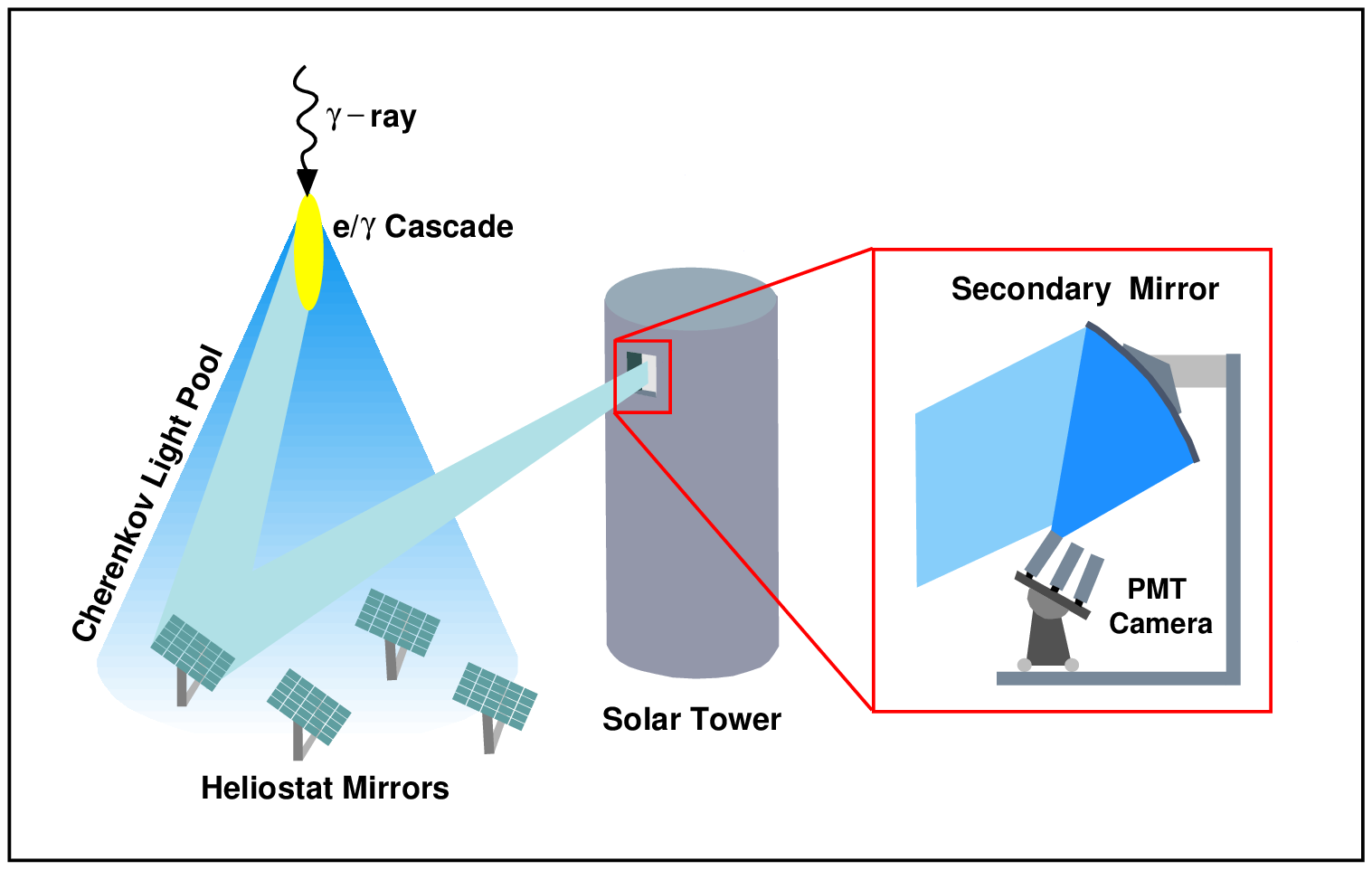}}
\caption{\label{concept} An illustration of the concept behind the 
STACEE gamma-ray detector. An incident gamma ray produces an air shower
and the particles therein generate Cherenkov light.
Heliostats on the ground direct this light towards a tower.
Secondary mirrors on the tower 
focus this light onto nearby photomultiplier tubes.
Each PMT views a unique heliostat.}
\end{figure}

STACEE makes use of the large steerable mirrors (heliostats) of the 
National Solar Thermal Test Facility (NSTTF) in Albuquerque, New Mexico.
The  NSTTF, built for solar power research,
has 224 heliostats, each with an area of 37 m$^2$, that can direct sunlight
onto a tower on the south side of the field.
STACEE uses 64 of these heliostats to sample the Cherenkov light pool 
and direct the light onto secondary mirrors located on the tower.
The secondaries focus the light onto cameras, each of which comprise a cluster of 
photomultiplier tubes (PMTs) such that each PMT views one heliostat. 
The process is illustrated schematically in Figure~\ref{concept}.

STACEE's efficiency for detecting optical photons is wavelength-dependent.
At wavelengths shorter than 350 nm 
the acceptance is cut off by the glass of the heliostat mirrors
which are aluminized on their rear surface. 
At longer wavelengths, the quantum efficiency of the PMTs drops to zero
between 600 and 700 nm. 
Thus STACEE's sensitivity is best between 400 and 500 nm and peaks at
about 420 nm.
Such a response is quite suitable for detecting Cherenkov light,
but it limits the range of possibilities for detection of 
interstellar laser pulses. 
The main concern is that extinction due to absorption and scattering
will be greater at shorter wavelengths ($A_\lambda/E(B-V)\simeq 3.5$
at $\lambda=420$ nm (Zagury 2001)) and might render a possible signal
too weak to see over a realistic distance. 
With the use of interstellar reddening maps ~\citep{schlegel} it can be shown that 
such an effect would amount to a loss of 10 to 20\% of the beam
intensity from sources in the directions of our targets.
However, it can still be argued that a longer wavelength would be
preferred, and we run the risk that this would be the choice of an
advanced civilization.

Pulses from each PMT are amplified and then split, one copy of which is sent to
trigger electronics, the other to an 8-bit 1 gigasample/s
flash-analog-to-digital converter (FADC).
The trigger copy is discriminated and dynamically delayed to account for the 
the observing geometry since arrival times of Cherenkov light on each
heliostat depend on the elevation and azimuth of the source being
tracked. 
Delayed pulses from the different channels are combined and a
minimum multiplicity is required within a coincidence window typically
12 ns wide. 
The FADC data are written to a buffer and 
when a trigger is generated, a 192 ns portion of this buffer is saved
to a data file.
Thus the data file consists of a series of events, each of which contain
64 192-ns traces. 

More details of the STACEE detector can be found elsewhere 
~\citep{chantell, hanna, gingrich}.

\section{Use of STACEE for OSETI} 

The potential of STACEE as a detector that is well suited to 
OSETI observations has been described elsewhere~\citep{covault}. 
Because of its large mirror area, STACEE is potentially more 
sensitive than detectors based on conventional
telescopes~\citep{howard,stone}.
Other atmospheric-Cherenkov 
detectors have already been used in a limited way for 
OSETI~\citep{holder} or are under consideration for future studies~\citep{armada}.

While it is true that, with 64 heliostats, each with an area of 37
m$^2$, STACEE has over 2300 m$^2$ of collection area for OSETI studies,
this has to be balanced with its large  
field-of-view (FOV) of about 0.6 degrees. 
This is a design feature, since sensitivity to Cherenkov light from air 
showers requires an FOV of about this size. 
This means that, though STACEE is sensitive to low-intensity laser
pulses, the air showers it detects provide a background 
that does not affect the optical-telescope detectors.
Thus, the many PMT channels of STACEE cannot be used independently to increase 
effective area; they must be used together to reject background from 
air showers generated by high-energy gamma rays and by (the far more copious) 
charged cosmic rays.

There are several criteria for rejecting  
backgrounds from air showers, the most powerful
of which are uniformity and multiplicity.
A laser pulse from a distant source will uniformly illuminate the
heliostat field, so we would expect every PMT channel in the STACEE 
detector to report the same number of photons, though with deviations
due to Poisson fluctuations and channel-to-channel efficiency differences.

\section{Observations}

\subsection{OSETI Data Runs}

The data presented here were acquired between January and May 2007.
During this time, STACEE was being run largely 
to complete a data set on the active galactic
nucleus 1ES 1218+304~\citep{muk1218}. 
During the observing hours when this source was not near transit, we kept the 
detector operational and staffed in order to be ready to follow up
gamma-ray burst (GRB) alerts.
We used these stand-by hours for a series of 10-minute OSETI 
observations on a list 
of candidate stars. 

\begin{figure}
\centerline{\includegraphics[width=1.0\textwidth]{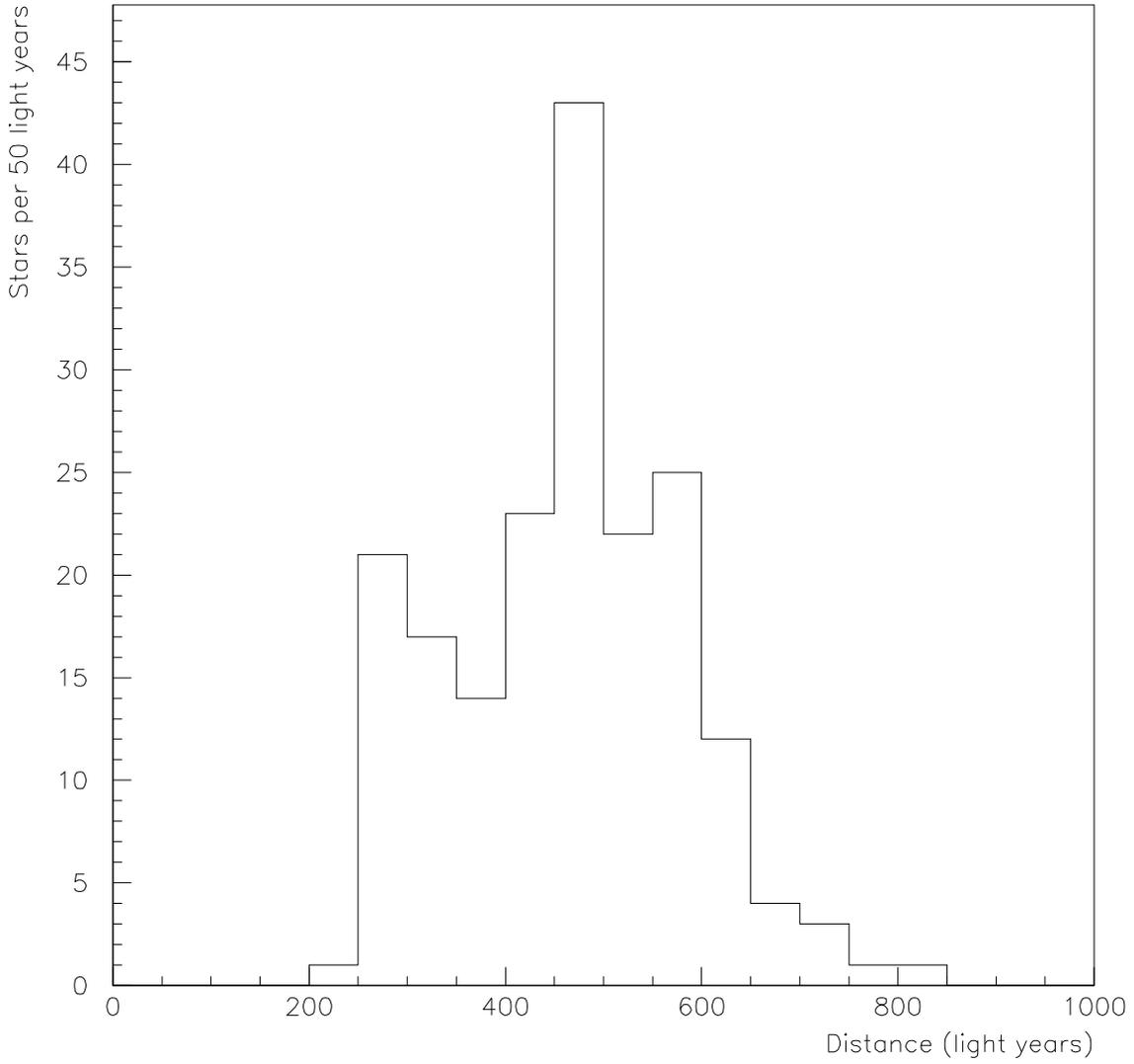}}
\caption{\label{distance} 
Histogram of the distances to the 187 stars 
observed during the STACEE OSETI campaign.}
\end{figure}

\begin{figure}
\centerline{\includegraphics[width=1.0\textwidth]{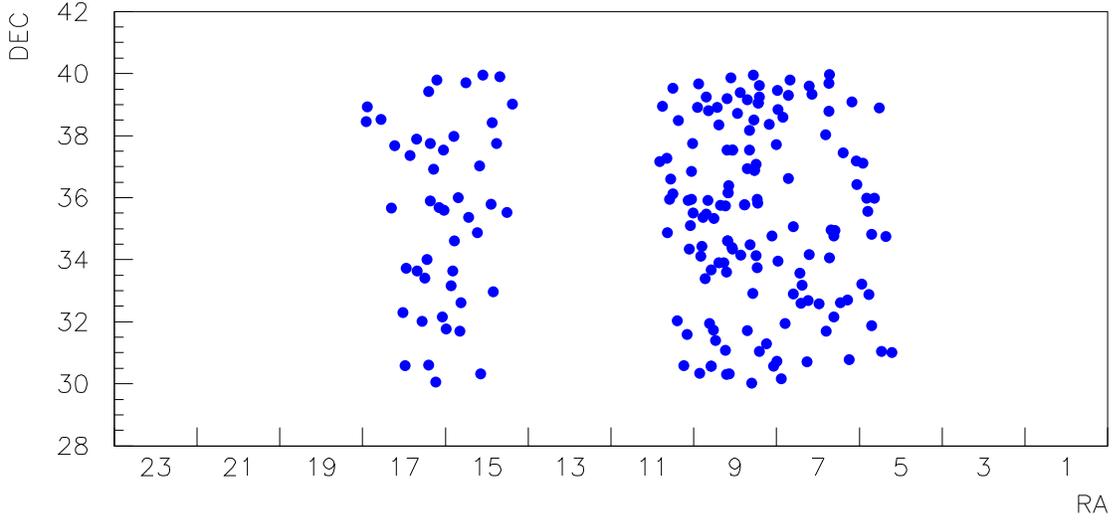}}
\caption{\label{radec} Declination (in degrees) vs right ascension (in
hours) for the stars observed during the STACEE OSETI campaign. The space
of possibilities was limited by the requirement that the stars transit close 
to zenith.}
\end{figure}

\begin{figure}
\centerline{\includegraphics[width=1.0\textwidth]{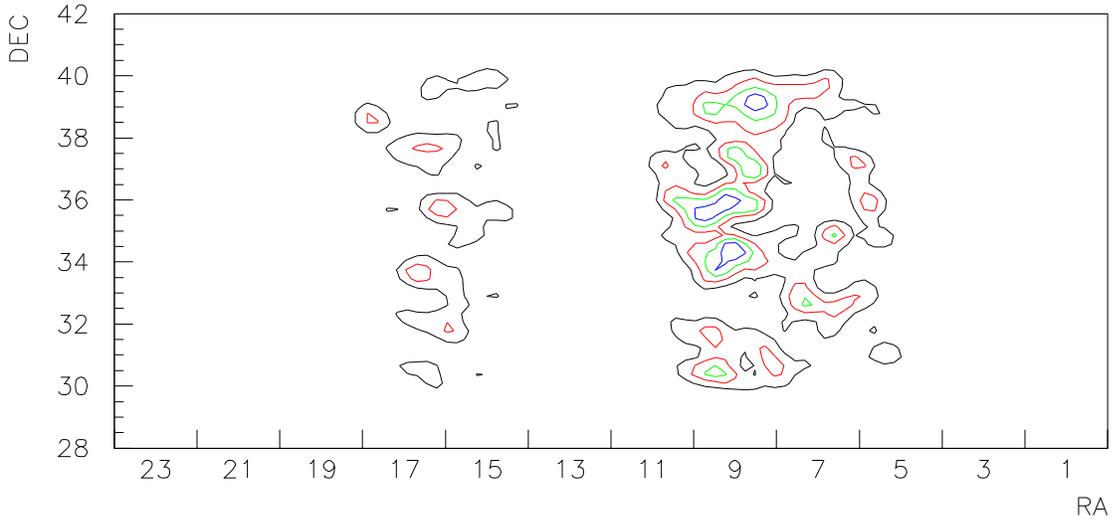}}
\caption{\label{contour} Contour plot of sensitivity as a function of
target direction. 
The positions of the stars plotted in Figure~\ref{radec} have been
smeared with a Gaussian function having a full-width corresponding to
the STACEE field-of-view (approximately 0.6 degrees). 
The contour step size is 20\% of the peak sensitivity.}
\end{figure}

We selected stars from the HabCat catalog~\citep{habcat}, 
a list of nearby Sun-like
stars that conceivably harbor habitable planets.
The targets chosen for STACEE were those that would be near zenith when they
transited, since the optics of STACEE are optimal for sources overhead.
All told, we observed 187 stars. A histogram of their distances from Earth 
is shown in Figure~\ref{distance}.
The celestial coordinates of the stars are plotted in Figure~\ref{radec}.
The Hipparcos catalog identifier of each star and the date on which it was observed is listed in Table~\ref{targets}.
We note that the FOV
of STACEE is approximately 0.6 degrees
so there is also sensitivity to any sources located in interstellar 
space in the region of the targeted stars.
(It has been argued~\citep{dyson} that advanced civilizations might
  inhabit ``Dyson Spheres'' which completely enclose their host stars
  and are therefore not associated with visible stars. This hypothesis
  was originally put forward in support of the notion of looking for
  extraterrestrials at infrared wavelengths. However it does not exclude
  the possiblility that such beings could be sending out laser pulses
  of the type we can detect.)
In regions between closely-spaced stars the effective sensitivity is
increased since that region of space is observed for a longer time. 
This is quantified in Figure~\ref{contour} where the points from 
Figure~\ref{radec} have been smeared with a Gaussian function with
full-width of 0.6 degrees.
The contours represent steps of 20\% of the peak sensitivity.

The data were acquired via the standard STACEE trigger system, which
was configured for GRB follow-up observations. 
We used this trigger for the OSETI runs to have the capability to switch 
quickly to GRB observations, with a minimum number of changes, following any 
alert from the network.
For OSETI running the discriminator threshold for each PMT was set at 90 mV, 
which corresponds to approximately 9 photoelectrons.

\subsection{Optical Throughput Measurements}

Almost all STACEE 
triggers result from air showers caused by charged cosmic rays
and a small number arise from gamma-ray events. 
In both cases, a pool of Cherenkov light is distributed over the heliostat 
field and this pool has a characteristic structure in space and time. 
Typically there will be systematic differences from channel to channel in the 
amount of light received and its time of arrival.
Thus a very powerful discriminant against air showers 
is the requirement that every PMT channel report a pulse size consistent
with the arrival of a uniform flux of optical photons.
To apply such a criterion it is necessary to unfold detector effects, two of
which are geometric acceptance and electronic gain.
The gains of the PMTs and their amplifiers are equalized with a 
laser calibration system~\citep{hanna_l} but the efficiencies of the optical 
components (heliostats, secondary mirrors and light concentrators) 
also need to be accounted for.
As described elsewhere~\citep{hanna}, the effective FOV of each heliostat depends on
its distance from the tower, and the channel-to-channel differences are 
roughly compensated for by using optical concentrators with different
angular acceptances attached to the PMTs.
Additionally, collection of light from the heliostats has slight 
channel-dependent variations.
For example, the PMT cameras occult some of the light from the heliostats, with
the exact amount depending on the particular heliostat.

\begin{figure}
\centerline{\includegraphics[width=1.0\textwidth]{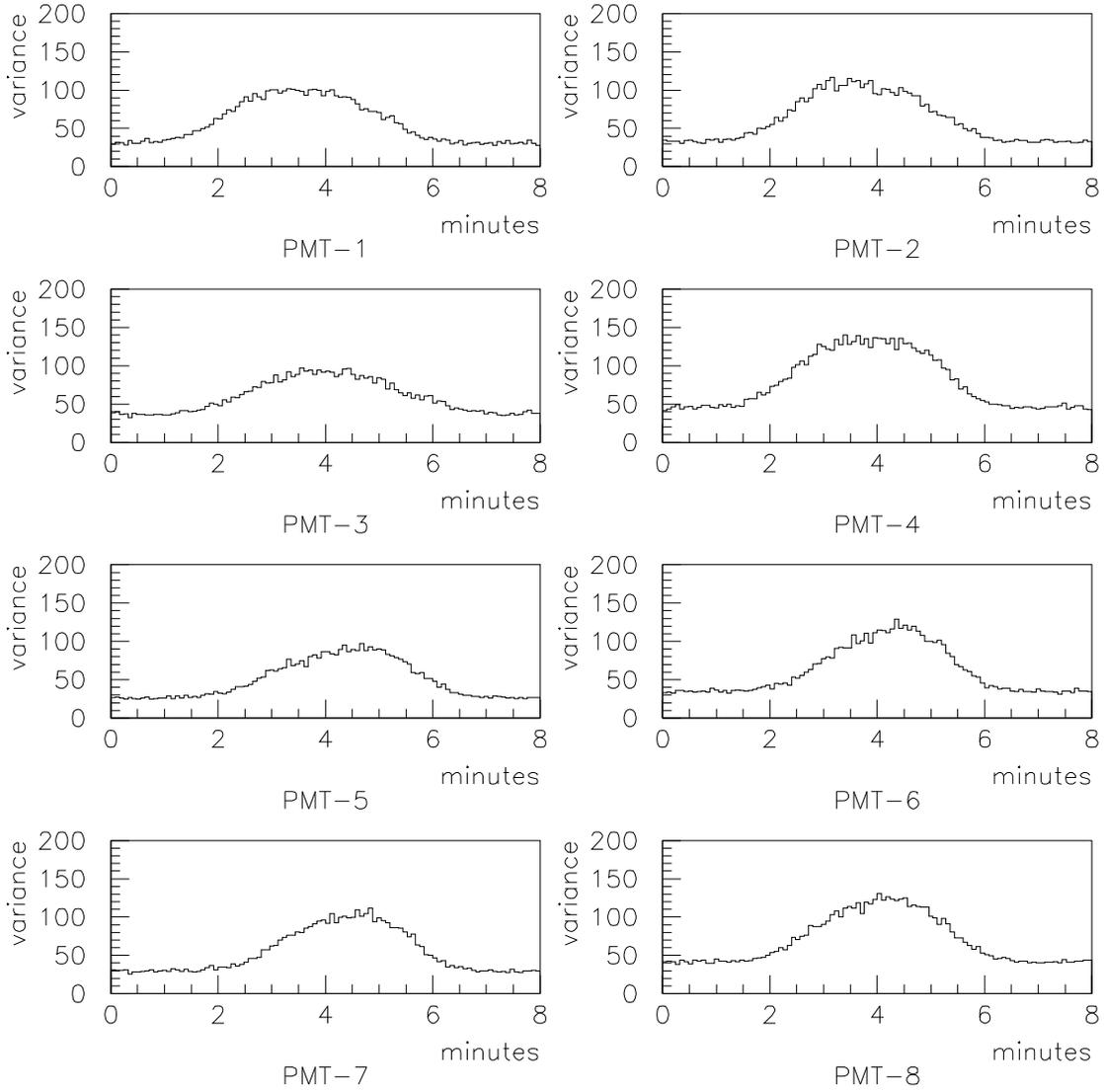}}
\caption{\label{varplot}
FADC baseline variances (in digital-counts-squared) vs time (in minutes)
for a drift scan with a bright star. Each plot corresponds to a different 
PMT channel. The deviation of the trace at 4 minutes
from the baseline value is a relative 
measure of the net throughput of the channel.}
\end{figure}

Fortunately, we can directly measure the net result 
of these effects by performing 
drift scans on bright stars.
We point the heliostats to a location that matches the declination 
of a star and is four  
minutes ahead of it in right ascension.
We disable heliostat tracking so that each mirror is held in a fixed
position.
We record photocurrents for eight minutes as 
the star drifts through the FOV of each PMT channel. 
Sample plots from the first eight PMT
channels are shown in Figure~\ref{varplot}.
In this figure we plot FADC baseline variance vs time. 
(The baseline variance is linearly related to photocurrent but its 
channel-to-channel calibration is more precisely known than that for the 
photocurrent.)
As the star drifts into the field-of-view, the current rises to a maximum and
then
returns to its quiescent value. 
The curve is usually asymmetric due to details of the geometry of the heliostat
position and the corresponding PMT.
 
We calculate the difference between the current at the four-minute mark 
and the average of the currents for the first and last 30 seconds of the scan
and use this quantity to define the throughput for the PMT channel. 
Note that in some cases the maximum deviation does not occur precisely at 
four minutes because of a slight mis-pointing of the heliostat. 
For the throughput calculation, however, it is 
important to use the value at four minutes since that takes into
account the effect of the slight mispointing, which will also be
present during OSETI runs.

This method of estimating net throughput is quite powerful. 
Since the photocurrent depends linearly on the number of photoelectrons 
arriving at the first dynode in the PMT, we automatically account for 
quantum efficiency and collection efficiency in the PMT, as well as gains
in the electron multiplier structure and in the downstream amplifiers. 
This is in addition to the geometric effects of light collection, reflection
and occultation.

\begin{figure}
\centerline{\includegraphics[width=1.0\textwidth]{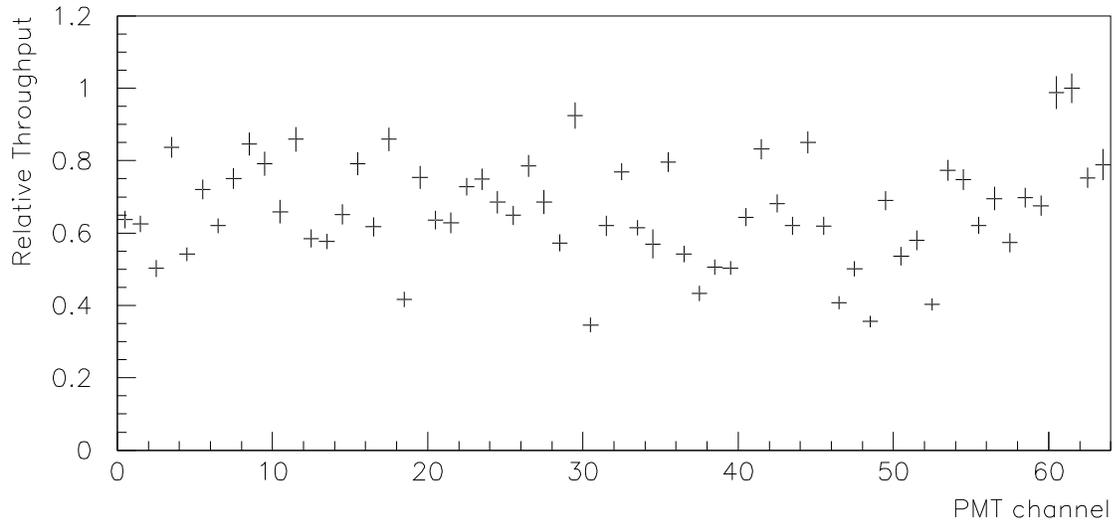}}
\caption{\label{throughput}
Relative throughputs vs PMT channel number as measured by the drift scan 
technique.}
\end{figure}

After calculating a throughput for each PMT 
channel, the values are scaled such 
that the channel with the largest throughput has unity value.
The relative throughputs are shown in Figure~\ref{throughput}. 
Most channels are within a factor of two of the most efficient one and all
are within a factor of three.

\subsection{Optical Throughput Calculations}

\begin{figure}
\centerline{\includegraphics[width=1.0\textwidth]{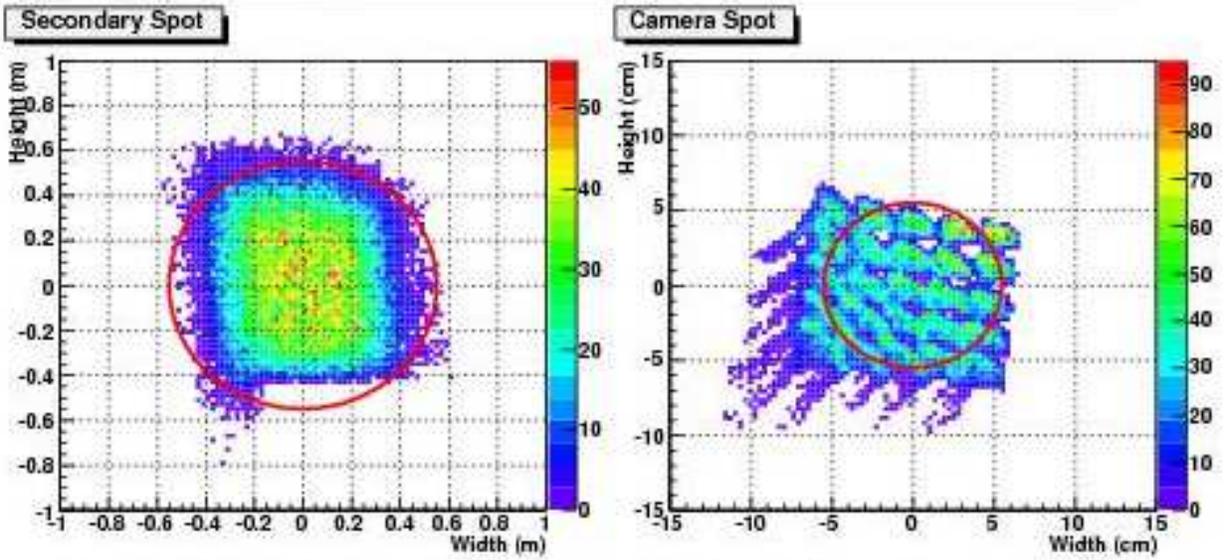}}
\caption{\label{raytrace}
Results for PMT channel 49 from the ray-tracing program used to 
evaluate absolute throughput.
Left panel: impact points of photons at the plane of the secondary mirror.
The 2-m circle indicates the position of the mirror. The blank area 
at the lower right is 
caused by occultation from the camera structure. 
Right panel: photon impact positions at the entrance to the optical 
concentrator on the PMT. Comatic aberration as well as structure due to the heliostat
facets can be seen.}
\end{figure}

The drift scans are well-suited for measuring relative response from
channel to channel, but they cannot be used to determine the absolute
optical throughput.
To do this we use a ray-tracing program that models all the optical components
using measured quantities such as reflectivities and transmissions. 
An indication of the detail of the ray-tracing program is given in 
Figure~\ref{raytrace}, which shows the results of tracing the trajectories of
photons that arrive vertically onto the heliostat field.
The left panel shows the impact points of photons 
at the secondary mirror on the tower. 
The outline of this mirror is shown as a circle and the blank 
region in the lower part is caused by occultation from the camera structure.
In the right panel, impact points at the focal plane are shown; the circle 
corresponds to the entrance aperture of the optical concentrator.
Aberration due to coma,  as well as the structure caused by the 25 facets
of the heliostat can be seen.

Using the ray-tracing program we calculate an average effective area per
heliostat of approximately 3 m$^2$ at a wavelength of 420 nm. 
The effective areas range from 1.5 m$^2$ to 3.5 m$^2$. 
At 1.5 m$^2$ a relatively large amount of light is lost due to occultation and 
the mis-match between the heliostat image size and the optical concentrator 
acceptance; at 3.5 m$^2$, such losses are less severe. 
As noted above, the nominal area of each heliostat is 37 m$^2$, so the overall
efficiency 
is less than 10\%. Most of this is due to the quantum efficiency of the PMTs.

\subsection{STACEE Sensitivity}

\begin{figure}
\centerline{\includegraphics[width=1.0\textwidth]{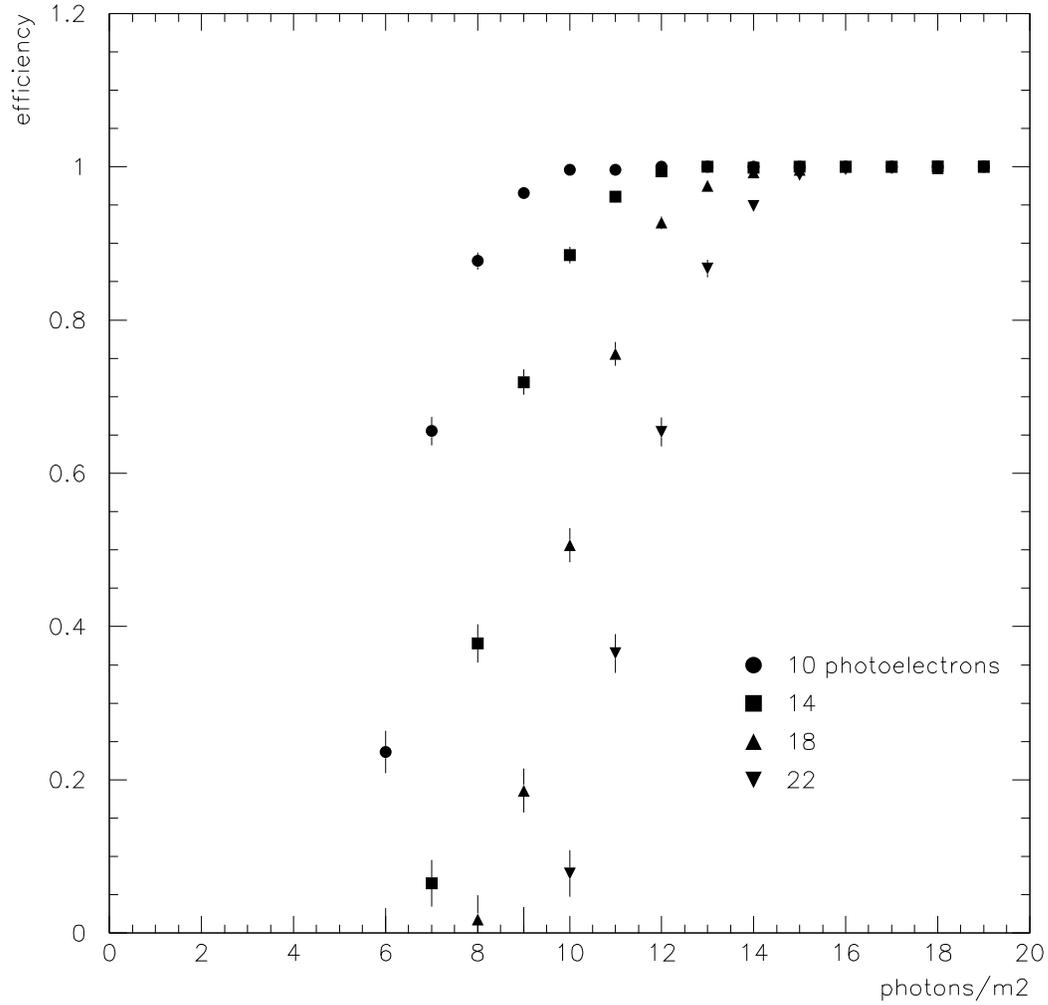}}
\caption{\label{accept_curve}
Efficiency vs optical flux levels for the STACEE detector for four threshold
values. An effective area of 3.5 m$^2$ for the most efficient PMT 
channel has been
assumed and all other channels have been scaled according to the relative 
throughputs plotted in Figure~\ref{throughput}.
Using a 10-photoelectron analysis threshold, we find no candidate laser flashes
in the STACEE data.
}
\end{figure}

To calculate the optical fluxes to which STACEE would be sensitive, we 
used a simple Monte Carlo program to simulate the response of our
detector to uniform light pulses.
Fluxes (in photons/m$^2$) were simulated over a range of values,  
and the mean numbers of photons 
expected in each PMT channel were calculated via the optical throughputs 
obtained from the drift scans. 
The PMT channel with the best throughput was assigned an effective area of 3.5
m$^2$, and the other channels were scaled accordingly.
The resulting mean values for the 
expected numbers of detected photons were used
to generate a characteristic Poisson distribution for each PMT channel.
64-channel events were generated by sampling these distributions and
the number of detected photons in each channel was corrected for the relative
acceptance of the channel.
Thus the mean corrected photoelectron count  
should be the same for all channels but the fluctuations 
depend on the relative acceptances of the PMT channels.
An event was accepted if all channels had a corrected photoelectron
count above threshold.
Results of this calculation are shown in Figure~\ref{accept_curve} where 
efficiency as a function of flux level is plotted for four different threshold
values.
For typical thresholds, the detector is sensitive to fluxes of the order of 
10-15 photons/m$^2$. 
This is very much a wavelength-dependent statement; these calculations were
carried out assuming a laser wavelength of 420 nm. 
For longer wavelengths our sensitivity is less, due to the fall-off in quantum 
efficiency of the PMTs.

\section{Data Analysis}

Each data run resulted in a file that 
contained approximately 1500 events, each of
which comprised 64 FADC traces, each 192 samples long.
The FADCs used an 8-bit digitizer so that each sample corresponded to a
number from 0 to 255.
A typical FADC trace consisted of a fluctuating baseline, with an average 
value of about 220 counts, with a negative-going pulse between sample
numbers 80 and 120.
Ancillary data such as the time of each event, as read from a GPS clock,
made up the rest of the data file.

The data analysis program was designed to re-apply the trigger criterion 
with tighter timing windows and select events with acceptance-corrected 
pulses which were consistent, allowing for Poisson fluctuations, with 
coming from a uniform flash of light.
The data processing was done in four stages. 

\begin{figure}
\vskip 4.0 cm
\centerline{\includegraphics[width=1.0\textwidth]{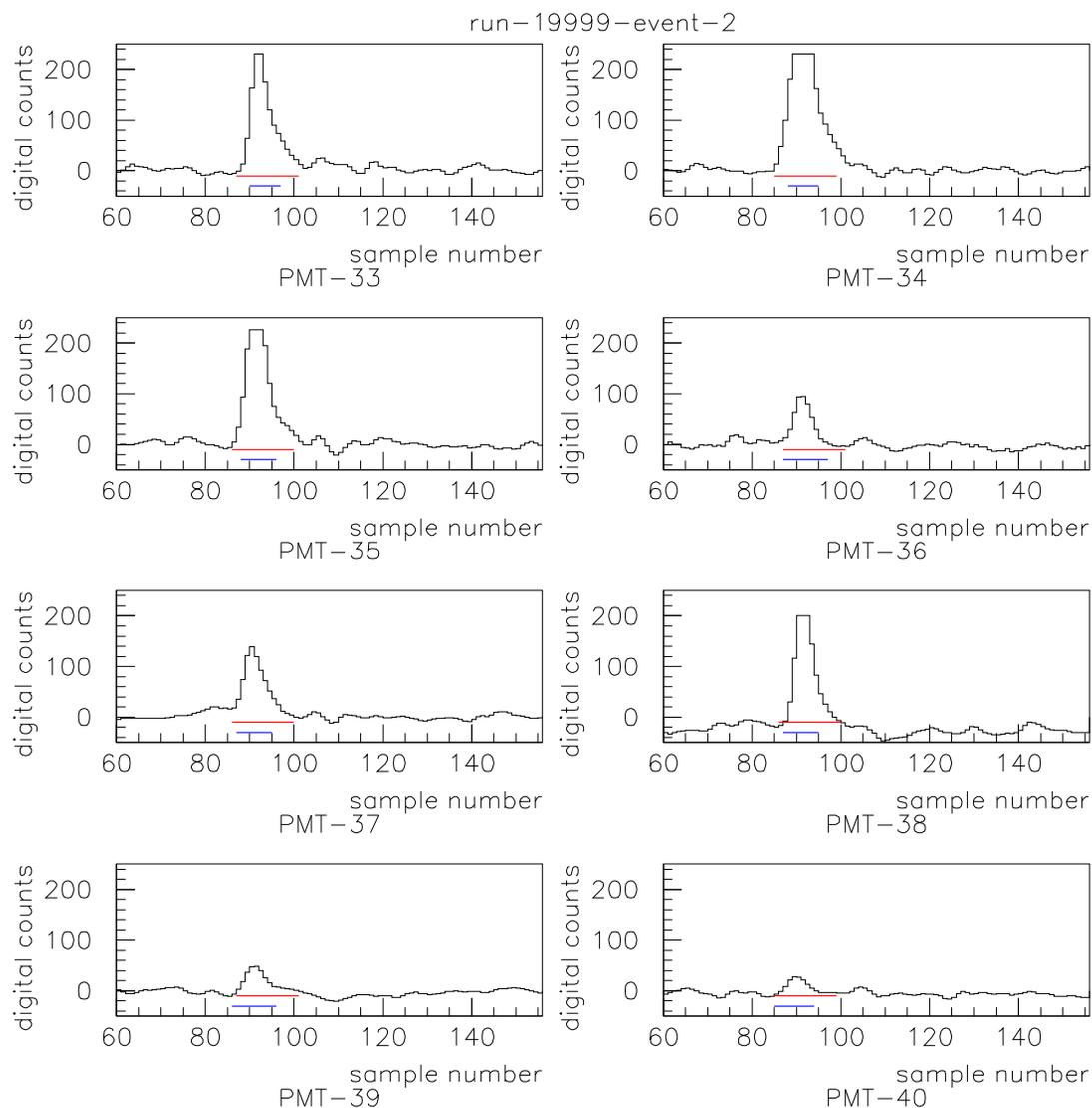}}
\vskip -4.0 cm
\caption{\label{sample_data} 
Sample traces from eight PMT channels of a particular event. 
The channel numbers are shown below each trace.
Samples 61 through 156 (of a possible 192 saved samples) are displayed.
The upper line shows the range over which the pulse is integrated to
give an estimate of the charge.
The lower line below indicates the window used in searching for a valid pulse.
Although the window for the pulse search can 
vary from channel to channel, the integration gate is fixed at 15 ns.
Note that the flat tops of the traces for channels 34 and 35 are 
due to saturation.}
\end{figure}

In the first stage, the first 50 FADC samples for each PMT channel
were histogrammed and fit to a Gaussian distribution.  
The mean was saved as the pedestal for the channel and the
root-mean-square standard deviation was saved to 
be used in defining a cut threshold.
The early part of the trace was used to avoid biases from in-time pulses. 
After the pedestal was computed, it was subtracted and the trace was inverted. 
(See see Figure~\ref{sample_data} for an example of typical traces.)
Any channel with at least one sample in its trace above a cut of 3
standard deviations 
(typically 15-20 counts) was deemed to be hit. 
If, in the raw trace, any samples were at zero (i.e., the pulse had 
``bottomed out'') the channel was labelled as saturated (indicating a
light level beyond the dynamic range of the digitizer).

In the second stage, hit but unsaturated PMT channels were used to calculate
an event time. Channels with maximum sample size more than 50 counts above 
baseline were used for timing, with the time of the maximum sample used
as the time of the pulse.
The average of these times from all available channels
was used as the event time; due to details of the 
trigger, this fluctuated by order 10 ns from event to event.
The deviation from event time for each channel was histogrammed and fit, 
at the end of the run, to a Gaussian to define a timing window 
(mean plus or minus 2.5 standard deviations) for use in the following stage.
The width of the timing window so defined was channel-dependent with most 
channels having a width of 10 ns, though some at the far edge of the 
heliostat field had windows that were 25 ns wide.
The use of narrow, channel-specific search windows enabled the use of the 
lowest possible thresholds in looking for valid pulses.
  
In the third stage, the event time and the window limits were used to define 
a search region for each PMT channel.
The maximum sample within the window was used to define the peak of the pulse
and the pulse charge was computed by summing the 15 samples starting 5 samples
before this point. 
The channel was counted as hit if the maximum sample was above the cut 
value defined above.
(Note that a channel defined as hit in the second stage of the analysis 
could lose this status at this point because the above-threshold sample in 
stage 2 was from outside the search window.)
This procedure produced a charge for all channels, even those 
that were below trigger threshold.

The analysis can be summarized while viewing Figure~\ref{sample_data}.
Displayed in this figure are traces from eight FADC channels for an 
arbitrary event. 
The central portion
of each trace (samples 61 through 156) is displayed in the figure.  
The upper line below each pulse delineates the 15 ns integration window
used for charge estimation. 
The lower line indicates the range over which samples are searched 
for a pulse.

For events with at least 55 PMT 
channels that have a sample above threshold in the 
search window, all 64 integrated charges were saved in a file for further 
processing.
Although the signature of a laser pulse would be 64 hit channels, the cut 
at 55 
allows for dead channels in some runs, as well as Poisson fluctuations on 
low-intensity candidates. 
 
For the selected events, charges were converted from units of 
counts-ns 
to photoelectrons with a scale factor of 20 counts-ns per photoelectron.
This factor was determined via special calibration and gain-equalizaton 
runs made with the laser calibration system~\citep{hanna_l}.
Gaussian ($\sqrt{N}$) uncertainties, which are statistically adequate
for $N > 10$, were assigned to the photoelectron 
estimates at this point. 
The photoelectron estimates and their uncertainties were multiplied by the 
relative throughput corrections (Figure~\ref{throughput}). 
Using only PMT channels that reported a corrected photoelectron count 
of at least 10, we computed an average number of photoelectrons for the event.
The threshold was applied to avoid including channels that, for some reason 
were not functioning properly during the run. 
(On some occasions one or two 
heliostats were stowed due to mechanical problems.) 
If an event passed further selection cuts, the low-charge
channels could be included in the calculations after checking on the hardware 
status. 
Typically, more than 50 channels contributed to the average photoelectron 
calculation. 

\begin{figure}[h]
\centerline{\includegraphics[width=0.7\textwidth]{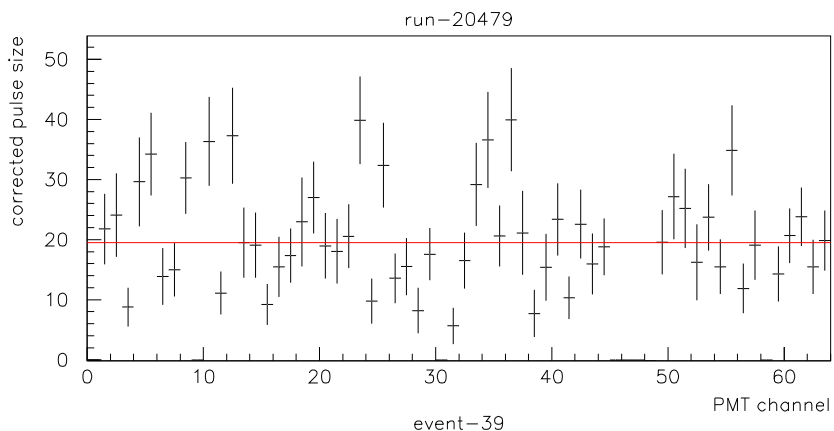}}
\vskip 0.5 cm
\caption{\label{passed-ex-pe}
Example of an event singled out for attention during the final stage 
of analysis.
The corrected pulse size (in photoelectrons) is plotted vs PMT 
channel number, together with the average computed using channels reporting 
more than 10 photoelectrons.
The $\chi^2_{DOF}$ for this event is 1.64.}
\end{figure}

\begin{figure}[h]
\centerline{\includegraphics[width=0.7\textwidth]{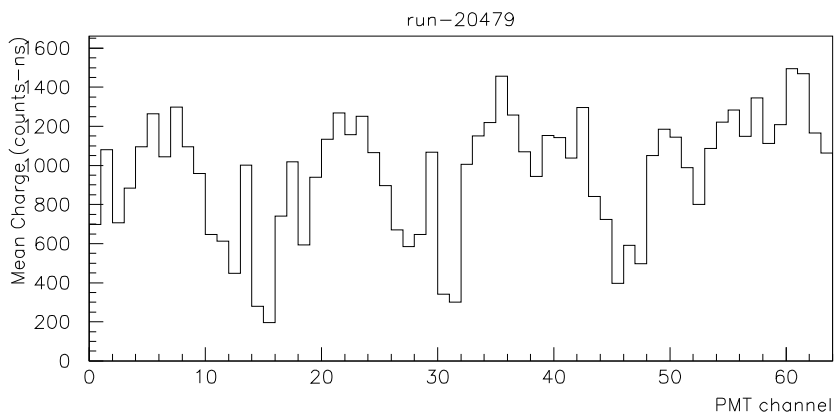}}
\vskip 0.5 cm
\caption{\label{passed-ex-means}
Mean charge vs PMT channel number for the run containing the candidate event 
shown in Figure~\ref{passed-ex-pe}. Units are counts-ns.
The structure is due to array geometry, 
trigger effects, and the distribution of light from air showers. The fact 
that each channel reports a non-zero mean value indicates that there were no 
defective channels in this run.}
\end{figure}

To search for events consistent with
uniform light illumination, we characterize the extent to which the
measured flux in all PMT channels is statistically consistent with a
single value.  
We define a standard chi-squared statistic to
characterize the extent to which channels deviate from the the average
value.  
The chi-square per degree-of-freedom ($\chi^2_{DOF}$) for such
a large number of statistically independent channels should be very
strongly peaked at a value of 1.0 for events that are consistent with
uniform illumination. 
Events with ($\chi^2_{DOF}$) significantly
greater than 1.0 are considered to be non-uniform, and are tagged as air
showers. 
For example, for n=50 the distribution expected for laser pulses 
is approximately
normal with a standard deviation of 0.3. 
Events with $\chi^2_{DOF} < 2.0$ were individually checked.
An example of data from such an event is shown in Figure~\ref{passed-ex-pe}.
The corrected number of photoelectrons for each of the 64 PMT 
channels is plotted
vs channel number, along with the fitted average.
Figure~\ref{passed-ex-means} shows the mean charge (in counts-ns) 
vs channel number for all events in the run.
The uneven 
structure is caused by the distribution of Cherenkov light in air showers 
and the efficiency for triggering on such showers as a function of where they 
hit the heliostat field. 
The fact that no channels reported zero for mean charge indicates that they 
were all in good working order for this run.
This result indicates that 
the zeros in PMT channels 45-48 in Figure~\ref{passed-ex-pe} are due to a
lack of light in the corresponding heliostats. This is 
not unexpected in air-shower events.

\begin{figure}[h]
\centerline{\includegraphics[width=0.7\textwidth]{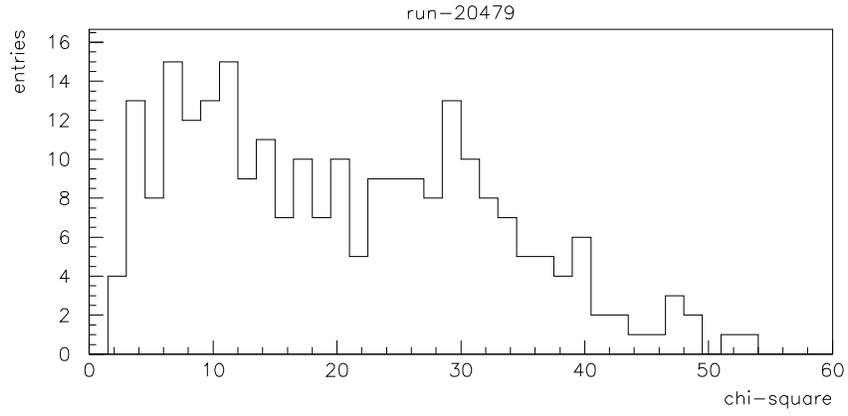}}
\vskip 0.5 cm
\caption{\label{passed-ex-chi2}
Distribution of $\chi^2_{DOF}$ for the events in the run containing the event
in Figure~\ref{passed-ex-pe}.
}
\end{figure}

The distribution of 
$\chi^2_{DOF}$ for all fits in the run is shown in Figure~\ref{passed-ex-chi2}. 
Most runs have one or two events that have $\chi^2_{DOF} < 2.0$
at this stage. 
These are invariably rejected because they have PMT channels that have 
below-threshold pulse sizes (often zero) even though the mean-charge
distributions indicate that they are in good working order.
These small pulses were left out of the initial average and 
$\chi^2_{DOF}$ calculation because of the 10-photoelectron threshold designed
to accomodate possibly bad channels.
When they are included, the $\chi^2_{DOF}$ gets much larger.

As can be seen in Figure~\ref{accept_curve}, with the 10-photoelectron
threshold used in the analysis, we were fully efficient for flashes
that resulted in flux levels above 10 photons/m$^2$ at the detector.
After processing all the data runs, we did not have any candidates consistent
with the arrival of pulses of at least this intensity.

\section{Discussion}

In the field of elementary particle physics, searches for new particles and
related phenomena are common.
Examples include the hunt for the weak interaction bosons (W and Z), the 
top quark and neutrino oscillations. 
These searches were ultimately successful.
Presently ongoing searches include the quest for axions, the Higgs boson, and
dark matter particles such as WIMPs.  
In such searches there is a space of possibilities with coordinates such 
as particle mass and interaction cross-section. 
Null results are presented as exclusion plots, which indicate what parts of the 
parameter space have been ruled out in that they do not warrant further consideration.
Progress is made by shifting the boundaries of these excluded regions into 
unexplored regions. 

OSETI searches are not so systematic.
It is possible to imagine repeating the same experiment some time after
it was first carried out and obtaining different results. 
A null result could be changed to a positive result in the event a distant 
civilization has, in the interim started sending out detectable signals. 
Repeating the experiment with a slight change in one aspect, e.g., dwell 
time on the target, might lead to a successful detection. 

What did we learn from the results of our search with the STACEE detector?
We can say that blue-green laser pulses of detectable intensity were not being 
sent, in our direction and from the vicinity of the target star, within those 
10-minute intervals during which our search was directed at them.
Saying more is difficult. 

Instrumentally, we have learned that use of a heliostat array for OSETI
searches is possible. 
However 
the large mirror area is reduced to a much smaller effective area by the
quantum efficiency of the PMTs and various geometric factors. 
The wavelength range is rather small and this reduces the amount of 
parameter space that can be investigated, though 
this shortcoming could be reduced with PMTs chosen for OSETI rather than
Cherenkov astronomy.

The large number of heliostats cannot be used to synthesize a giant collector,
and, thereby, increase sensitivity to faint laser pulses. 
This is because multiple channels are needed to reject the huge background 
of nanosecond optical flashes caused by extensive air showers.
The simple analysis described in this paper requires a larger photon
flux than the 2 photons/m$^2$ foreseen by Covault (2001) but it is not
excluded that, using techniques described in that paper (fitting the
times to a wavefront, cutting on arrival direction, {\it etc.}), one could
lower the required flux.
  
The loss of effective collection area is offset somewhat by the
larger field-of-view afforded by an instrument of this type.
We have not explored how few channels can be used to suppress 
Cherenkov background sufficiently, but it 
could be possible to divide the detector into several smaller detectors 
to allow for several regions to be monitored at once
This would require more electronics to construct the extra triggers.

How does the STACEE search compare to other OSETI efforts?

The Harvard-Princeton search ~\citep{howard} was much more extensive,
with nearly 16,000 observations on over 6000 objects 
totalling 2400 hours of data. 
The typical time per target was 24 minutes. 
The stated sensitivity was 100 photons/m$^2$ in a window of 5 ns in
the 450-650 nm band for the Harvard telescope.
The Princeton telescope, which observed in parallel for 1142 of the
targets, had a similar sensitivity (80 photons/m$^2$).

A team at the Lick Observatory made 5000 targeted observations with use 
of a conventional optical telescope~\citep{stone}. 
Their flux sensitivity was similar, but their time-on-target was 10 minutes.

The Harvard all-sky-detector team~\citep{howard07} reported that they achieved
a sensitivity of 95 photons/m$^2$ in a window of 3 ns.
They ran in a drift-scan mode with a consequent one-minute dwell time
per source point.

A group ~\citep{holder} searching for
OSETI signals in archival data from the Whipple
10-meter gamma-ray telescope, another Cherenkov-based instrument,  
demonstrated that a sensitivity of 10
photons/m$^2$ is possible.
Each run in their search had a length of 28 minutes, but the same part
of the sky was targeted in many different runs since gamma-ray
observations typically require very long exposures.
Because of this the total time spent on a given target was very large.

One conclusion from this comparison is that 
the large mirror areas of the Cherenkov
telescopes can be used to improve flux sensitivity by about one order of
magnitude compared with the smaller optical telescopes.
To put the flux of 10 photons/m$^2$ in perspective, we consider a 400 nm
laser at a distance of 1000 light years. 
A 0.3 MJ pulse would contain $6 \times 10^{23}$ photons, and if these
were formed into a diffraction-limited beam with a 10 m mirror, the
beam diameter at the Earth would be $4 \times 10^{11}$ m. 
Assuming that the profile is such that half the photons fall within a
diameter of $2 \times 10^{11}$ m, an average flux density of 10
photons/m$^2$ is obtained.
Lasers with an order of magnitude greater pulse energy have already
been developed, so our search is not limited by what might
be considered reasonable technical achievements of distant civilisations.

\section{Future Progress}

The STACEE detector was dismantled during the summer of 2007 so there
will be no further observations of the type reported here.
Some members of the STACEE collaboration, however, have been studying
the possibilities for future detectors dedicated to OSETI. 
One idea that has emerged is the concept of multiple, distributed
OSETI detectors.

This idea foresees the use of a large number of inexpensive detectors,
each of which carry out drift scans in the manner of the Harvard all-sky
detector~\citep{howard07}. 
A detector element would consist of two pairs of telescopes, where each
telescope comprises a simple spherical mirror, approximately one meter in
diameter, with a single PMT at the focus.
The two members of a pair, both pointing at the zenith,
would be located on the order of 100 meters apart and operated in
coincidence mode with cables and simple electronics.
With each local coincidence, digitized pulses from the two PMTs would
be stored along with a GPS time stamp.
Most of these recorded events would be due to Cherenkov light from 
large air-showers, but the rate would not be large due to the
separation of the two members of the pair.
A second, identical pair directed at the same spot on the sky  
would be deployed several kilometers away such that coincidences
between the two pairs could not be caused by air-showers.
Data from the two pairs would be compared off-line using the GPS
time stamps to identify coincidences and the digitized pulses to check
for uniform illumination.

A large number of such installations could be built, possibly in
the style of the various outreach projects involving cosmic-ray
detectors located at high schools~\citep{carlson}.
With a large number of detectors, each scanning the same strip of sky
night after night, one would address what we consider to be one of the
most important limitations of the OSETI searches carried out to date:
duty factor. 
It is important to be live and on-target when a distant laser is
pointed our way. 
Sensitivity to faint laser pulses would  not be useful if the laser and
detector were not co-aligned at the crucial time.
A world-wide grid with spacing on the order of the detector
field-of-view would be useful and affordable, given the simple nature
of the detectors, which involve very basic electronics and no moving
parts.

Of course, the success of an OSETI campaign would depend on whether
civilizations exist that have the impetus to transmit laser pulses.
The number of such civilizations cannot be estimated reliably, though limits
can be set by using tools such as the Drake equation~\citep{drake}.
This equation attempts to identify the various factors needed
to make such an estimate, so that some of the
uncertainties involved can be quantified. 
Hetesi and Reg\'aly~\citep{hetesi} purported that the Drake equation
is not complete in that it neglects certain key factors, e.g.,
advanced civilisations may choose to
remain silent.
There are many good reasons to suspect that such civilisations would
be in the majority.

\section{Conclusion}

We have used the STACEE high-energy gamma-ray detector to look for fast 
blue-green laser pulses from the vicinity of 187 stars. 
These stars have been chosen because their characteristics are such that 
they may harbor habitable planets and they are
relatively close to Earth.
Each star was observed for 10 minutes, and we found no evidence
for laser pulses with intensities greater than about 10 photons/m$^2$
at 420 nm in any of the data sets.

\section{Acknowledgements}

We are grateful to the staff at the National Thermal Solar Test Facility 
for their enthusiastic and professional support. 
The STACEE project was funded in part by the U.S. National Science Foundation, 
the University of California, Los Angeles,
the Natural 
Sciences and Engineering Research Council, le Fonds Qu\'ebecois de la 
Recherche sur la Nature et les Technologies,
the Research Corporation, and the 
California Space Institute.




\clearpage

%% the file, uncomment the \plotone, \plottwo, and
%% If you need a layout that cannot be achieved with \plotone or
%% \plottwo, you can invoke the graphicx package directly with the
%% \includegraphics command or use \plotfiddle. For more information,
%% please see the tutorial on "Using Electronic Art with AASTeX" in the
%% documentation section at the AASTeX Web site,
%% http://www.journals.uchicago.edu/AAS/AASTeX.
%%
%% The examples below also include sample markup for submission of
%% supplemental electronic materials. As always, be sure to check
%% the instructions to authors for the journal you are submitting to
%% for specific submissions guidelines as they vary from

%% This example uses \plotone to include an EPS file scaled to
%% 80% of its natural size with \epsscale. Its caption
%% has been written to indicate that additional figure parts will be
%% available in the electronic journal.

\newpage

\begin{deluxetable}{l    l    l    l    l    l    l    l}
\tablecaption{Target Stars: the identifier refers to the Hipparcos
  Catalog and the Observation date is month and day in 2007.\label{targets}}
\tablehead{
\colhead{Identifier} & \colhead{Observed} &
\colhead{Identifier} & \colhead{Observed} &
\colhead{Identifier} & \colhead{Observed} &
\colhead{Identifier} & \colhead{Observed} 
}
\startdata
 41998 & 0124 & 36053 & 0217 & 47993 & 0222 & 76859 & 0317  \\		
 42686 & 0124 & 37374 & 0217 & 49770 & 0222 & 78751 & 0317  \\		
 41930 & 0124 & 38995 & 0217 & 71753 & 0223 & 79538 & 0317  \\		
 45334 & 0124 & 39112 & 0217 & 72892 & 0223 & 80206 & 0317  \\		
 47313 & 0124 & 39459 & 0217 & 72175 & 0223 & 81147 & 0317  \\		
 46033 & 0125 & 41530 & 0217 & 74260 & 0223 & 82432 & 0317  \\		
 47016 & 0125 & 42445 & 0217 & 44502 & 0312 & 36952 & 0318  \\		
 43596 & 0125 & 42763 & 0217 & 45883 & 0312 & 37991 & 0318  \\		
 43895 & 0125 & 42405 & 0217 & 46707 & 0312 & 38950 & 0318  \\		
 45126 & 0125 & 45175 & 0217 & 48202 & 0312 & 41382 & 0318  \\		
 45307 & 0125 & 46446 & 0217 & 48465 & 0312 & 41888 & 0318  \\		
 32304 & 0213 & 25614 & 0218 & 49283 & 0312 & 43024 & 0318  \\		
 34809 & 0213 & 26863 & 0218 & 50920 & 0312 & 44502 & 0318  \\		
 34977 & 0213 & 27361 & 0218 & 49241 & 0312 & 45053 & 0318  \\		
 36904 & 0213 & 28102 & 0218 & 51454 & 0312 & 45006 & 0318  \\		
 38480 & 0213 & 29617 & 0218 & 52885 & 0312 & 46993 & 0318  \\		
 26506 & 0216 & 30391 & 0218 & 44646 & 0314 & 72628 & 0319  \\		
 28005 & 0216 & 32168 & 0218 & 48566 & 0314 & 73829 & 0319  \\		
 27234 & 0216 & 32239 & 0218 & 49157 & 0314 & 74512 & 0319  \\		
 28707 & 0216 & 32211 & 0218 & 51804 & 0314 & 75579 & 0319  \\		
 29319 & 0216 & 33557 & 0218 & 52601 & 0314 & 76525 & 0319  \\		
 29874 & 0216 & 35193 & 0218 & 52119 & 0314 & 77402 & 0319  \\		
 31619 & 0216 & 34812 & 0218 & 41382 & 0316 & 78638 & 0319  \\		
 31431 & 0216 & 35825 & 0218 & 41642 & 0316 & 80326 & 0319  \\		
 34468 & 0216 & 38295 & 0218 & 43471 & 0316 & 72760 & 0320  \\		
 35962 & 0216 & 38949 & 0218 & 45006 & 0316 & 74156 & 0320  \\		
 37620 & 0216 & 39667 & 0218 & 45162 & 0316 & 75908 & 0320  \\		
 37589 & 0216 & 41292 & 0218 & 46993 & 0316 & 76656 & 0320  \\		
 39031 & 0216 & 42386 & 0218 & 48387 & 0316 & 78277 & 0320  \\		
 40037 & 0216 & 41270 & 0218 & 49529 & 0316 & 78559 & 0320  \\		
 41272 & 0216 & 42730 & 0218 & 49687 & 0316 & 80807 & 0320  \\	
 41521 & 0216 & 44525 & 0218 & 49111 & 0316 & 77469 & 0416  \\		
 42104 & 0216 & 45484 & 0218 & 50767 & 0316 & 77670 & 0416  \\		
 41691 & 0216 & 46314 & 0218 & 52052 & 0316 & 79381 & 0416  \\		
 44987 & 0216 & 47413 & 0218 & 77304 & 0316 & 80571 & 0416  \\		
 44559 & 0216 & 40398 & 0221 & 79105 & 0316 & 80128 & 0416  \\		
 24295 & 0217 & 41454 & 0221 & 79824 & 0316 & 81677 & 0416  \\		
 25030 & 0217 & 41888 & 0221 & 80417 & 0316 & 83075 & 0416  \\		
 25852 & 0217 & 42182 & 0221 & 81776 & 0316 & 84283 & 0416  \\		
 26849 & 0217 & 43024 & 0221 & 82938 & 0316 & 83326 & 0416  \\		
 27501 & 0217 & 44856 & 0221 & 45196 & 0317 & 84688 & 0416  \\		
 28727 & 0217 & 45053 & 0221 & 47739 & 0317 & 85879 & 0416  \\		
 30751 & 0217 & 47581 & 0221 & 48081 & 0317 & 87526 & 0416  \\		
 31972 & 0217 & 47191 & 0221 & 50185 & 0317 & 70297 & 0420  \\		
 31576 & 0217 & 46636 & 0222 & 49358 & 0317 & 70930 & 0420  \\		
 32658 & 0217 & 46022 & 0222 & 51460 & 0317 & 87680 & 0420  \\ 	
 32568 & 0217 & 47575 & 0222 & 51687 & 0317 \\
\enddata
\end{deluxetable}

%% If you are not including electonic art with your submission, you may
%% mark up your captions using the \figcaption command. See the
%% No more than seven \figcaption commands are allowed per page,
%% so if you have more than seven captions, insert a \clearpage

%% Tables should be submitted one per page, so put a \clearpage before

%% Two options are available to the author for producing tables:  the
%% deluxetable environment provided by the AASTeX package or the LaTeX
%% table environment.  Use of deluxetable is preferred.
%%

\end{document}